\documentclass[preprint]{elsarticle}

\newif\ifextended
\extendedfalse

\makeatletter
\let\@afterindenttrue\@afterindentfalse
\makeatother

\usepackage{subfig}
\usepackage{wrapfig}
\usepackage[T1]{fontenc}
\usepackage[utf8]{inputenc}
\usepackage{fancyvrb}
\usepackage{multibib}
\usepackage{graphicx}
\usepackage[dvipsnames]{xcolor}
\usepackage{siunitx}
\usepackage{xspace}
\usepackage[inline]{enumitem}
\usepackage{url}
\usepackage{booktabs}
\usepackage{caption}  
\usepackage[framemethod=TikZ]{mdframed}
\usepackage[bookmarks=false,hidelinks]{hyperref}
\usepackage{threeparttable}
\usepackage{java}
\usepackage{adjustbox}
\usepackage{float}

\usepackage[colorinlistoftodos,textwidth=10mm,textsize=tiny,obeyFinal]{todonotes}

\newcommand{\MyBox}[1]{\vspace{2mm}\noindent\framebox[\columnwidth][c]{\parbox[b]{0.95\columnwidth}{ #1 }}\vspace{2mm}}
\newcommand{\handora}{\textsf{Handora}\xspace}

\newcommand{\cdd}{Cognitive-Driven Development}
\newcommand{\xxx}{\textbf{\Huge{XXX}}}

\definecolor{shadecolor}{rgb}{0.75, 0.75, 0.75}

\usepackage{lstautogobble}  
\usepackage{color}          
\usepackage{zi4}            

\definecolor{bluekeywords}{rgb}{0.13, 0.13, 1}
\definecolor{greencomments}{rgb}{0, 0.5, 0}
\definecolor{redstrings}{rgb}{0.9, 0, 0}
\definecolor{graynumbers}{rgb}{0.5, 0.5, 0.5}

\usepackage{listings}
\lstdefinelanguage{diff}{
    morecomment=[f][\color{blue}]{@@},     
    morecomment=[f][\color{red}]-,         
    morecomment=[f][\color{blue}]+,       
    morecomment=[f][\color{red}]{---}, 
    morecomment=[f][\color{blue}]{+++},
}

\usepackage{balance}

\graphicspath{ {images/} }

\newmdenv [ %
 skipabove=\topsep,
 skipbelow=\topsep,
 leftmargin       = 0.2              ,
 rightmargin      = 0.2              ,
 splittopskip     = \topskip      ]{mh}

\begin{document}

\title{Cognitive-Driven Development Helps Software Teams to Keep Code Units Under the Limit!}

\author[1]{Gustavo Pinto}
\ead{gustavo.pinto@zup.com.br}

\author[1]{Alberto de Souza}
\ead{alberto.tavares@zup.com.br}

\address[1]{Zup Innovation, Brazil}

\begin{abstract}
Software design techniques are key elements in the process of designing good software. Over the years, a large number of design techniques have been proposed by both researchers and practitioners. Unfortunately, despite their uniqueness, it is not uncommon to find software products that make subpar design decisions, leading to design degradation challenges. One potential reason for this behavior is that developers do not have a clear vision of how much a code unit could grow; without this vision, a code unit can grow endlessly, even when developers are equipped with an arsenal of design practices.
Different than other design techniques, Cognitive Driven Development (CDD for short) focuses on 1) defining and 2) limiting the number of coding elements that developers could use at a given code unit.

In this paper, we report on the experiences of a software development team using CDD for building from scratch a learning management tool at Zup Innovation, a Brazilian tech company. By curating commit traces left in the repositories, combined with the developers' perception, we organized a set of findings and lessons that could be useful for those interested in adopting CDD. For instance, we noticed that by using CDD, despite the evolution of the product, developers were able to keep the code units under a small amount of size (in terms of lines of code). Furthermore, although limiting the complexity is at the heart of CDD, we also discovered that developers tend to relax this notion of limit so that they can cope with the different complexities of the software. Still, we noticed that CDD could also influence testing practices; limiting the code units' size makes testing easier to perform.
\end{abstract}
\maketitle

\section{Introduction}

The software development community has long recognized the importance of well-designed and modularized code to ease the maintainability and evolution process of a software product~\cite{Dewayne:Book}. From the works of David Parnas in the 70s, indicating the need for better approaches to support software aging~\cite{parnas1994software,parnas1987active}, there has been a significant number of design metrics~\cite{chidamber1994metrics,zage1993evaluating,shaik2010metrics}, tools~\cite{lefever2021lack,Tushar:Designite,Marcilio:JSS:2020}, and processes~\cite{uchoa2020does} that the research community envisioned to help software engineers in designing better software. Practitioners have also been fruitful in proposing well-adopted design approaches, such as clean code~\cite{martin2009clean}, open-closed principles~\cite{martin1996open}, etc. 

Generally speaking, these design techniques aim to help developers translate requirements into optimized code units\footnote{For the context of this work, a code unit is a source code file. We use the terms ``code units'' and ``classes'' interchangeably.}, given a set of constraints (imposed by the problem and the designer)~\cite{yau1986survey}. The research community has long recognized good design techniques as a critical factor in building reliable and understandable software~\cite{parnas1994software,yau1986survey}. Moreover, some of these design methodologies became so widespread that even novice programmers must understand some of them when applying for jobs~\cite{Matthias:ESEM:2022}. 

Intriguingly, despite the growing number of design practices that aid the development of well-designed software, developers still employ subpar design decisions~\cite{lavallee2015good}. In the current landscape of software development, it is not uncommon to find software products that suffer from design degradation~\cite{le2018empirical}, a scenario in which it becomes increasingly difficult to maintain the codebase while decreasing testability and reuse, impacting the overall effort to deliver new features~\cite{Leonardo:ICSE:2018,Oizumi:ISSRE:2019}.

One potential reason these design approaches are not yet killer tools in the programmer arena is that they are \emph{subjective}. Take, for instance, the catalog of bad design practices for refactoring testing code~\cite{van2001refactoring}. Although this catalog guides developers on what they could avoid/refactor, some of the items in this catalog are ultimately subject to one's interpretation. For instance, although assertions are undoubtedly valuable for detecting bugs, there is still little consensus on how assertions should be designed or documented (i.e., the assertion roulette bad smell~\cite{van2001refactoring})~\cite{Yucheng:FSE:2015,Kochhar:SEIP:2019,Aniche:CSMR:2013}. 

This limitation makes it difficult for researchers and tool builders to develop automated tools that find these bad practices with good precision. 
Another limitation is that these design practices do not provide a clear limit to developers. For example, how many smells a code unit might tolerate? What is the limit of design degradation we could handle before refactoring the bad smells? How many methods might a class have? How long should a testing method be? 

Without a clear understanding of this limit, code units might grow endlessly, making it harder for developers to reason about big code chunks. Indeed, recent research suggests that component size is a primary indicator of maintenance effort~\cite{Yossi:EMSE:2017}.
As pointed out by Kent Beck, one of the first agile practitioners and strong advocates for well-designed code, ``\emph{The goal of software design is to create chunks or slices that \textbf{fit into a human mind}. The software keeps growing, but the human mind maxes out, so \textbf{we have to keep chunking and slicing differently} if we want to keep making changes} (emphasis ours).''\footnote{\url{https://twitter.com/kentbeck/status/1354418068869398538}}.

Cognitive-Driven Development (CDD for short)~\cite{CDD:ESEM:2022,Souza:ICSME:2020,Pinto:ICEIS:2022} is a design coding technique that aims to reduce the complexity of code units (e.g., a class) by systematically posing a limit in the number of coding items --- that adds complexity to that code unit --- that could be used at once. Since CDD can be straightforwardly measured (in essence, it relies on counting the occurrences of complexity items in a given code unit), developers have little doubts about when and how they should apply it. Moreover, by limiting the number of items of complexity, CDD can guide developers to refactor code by increasing the awareness of classes (or methods) with higher complexity. \emph{Or so we think.}

\vspace{0.2cm}
\noindent
\textbf{\emph{This paper~~}} We present the first ``in vivo'' study on the use of CDD in an industrial setting. This study was conducted at Zup Innovation, a Brazilian Tech company. We have studied the use of the CDD practice for over one year.  
This paper shares details about the studied project, the teams, and how they used CDD.
Our goal with this report is to reflect on the team's (good and bad) experiences in using CDD on a daily basis to create a real-world software product. Therefore, our main research question is the following:

\MyBox{\textbf{RQ.} To what extent \cdd impact the size of code units?}

To provide answers to this question, we revisited the artifacts created during the software development process to achieve this goal. In particular, 
1) we mined our \texttt{git} logs, to find events that could indicate the use of the CDD practice, 
2) we watched our video calls, in which the team reflected on the pros/cons of using CDD, 
and 4) we studied the team's documentation, in which they shared their rationale for their design and architectural decisions. After curating this data, we presented our early findings to the development team to double-check whether our results are aligned with their perceptions. 

By reflecting on their experiences and analyzing the artifacts produced this year, we also contribute with lessons that could be useful for other teams interested in adopting CDD as their design technique. In summary, this paper provides the following contributions.

\begin{enumerate}
    \item A detailed description about CDD and a step-by-step guide that interested developers can follow to use it;
    \item A one-year reflection on the use of CDD as the main design practice for building an online learning management system at Zup Innovation;
    \item A set of lessons learned that could be useful for other software producing teams interested in using CDD as part of their design methodologies.
\end{enumerate}

\noindent
\textbf{Artifacts availability.} The data analyzed in this work are publicly available~\cite{cdd-artifacts}.

\section{What is CDD?}\label{sec:cdd}

CDD is a coding design technique that aims to reduce the cognitive load developers may face by limiting the number of programming constructs they can use in a given code unit. In this section we distill our rationale for building CDD (Section~\ref{sec:cdd:theory}), we introduce a guide that could help software producing teams interested in using CDD (Section~\ref{sec:cdd:practice}), and we highlight CDD's key characteristics (Section~\ref{sec:cdd:characteristics}).

\subsection{CDD Theory}\label{sec:cdd:theory}

CDD has its roots in the ``\textit{Magical Number Seven}'' theory~\cite{miller1956magical}, which is a well-adopted psychological theory that suggests that there is a \textbf{limit} (and often small) in the number of information items that could be processed in the working memory at a given time. If a large number of information items are provided to process some task, the short-term memory may become overloaded --- with an excessive cognitive load --- potentially hindering one's understanding. Experimental studies suggested that humans generally hold only seven (plus or minus two) information items in short-term memory~\cite{miller1956magical}. 

Another theory that CDD is built upon is the Cognitive Load Theory (CLT), proposed by Sweller~\cite{sweller1988cognitive}. CLT explains that any material has its \textbf{intrinsic complexity}, which varies according to the amount and arrangement of the information items that compose it. According to researchers~\cite{paas2004cognitive}, knowing the number of information items and their intrinsic complexities is crucial to guide instructors in presenting information at a pace and complexity level that the learners can understand. 



For CDD, we borrow the notions of \textbf{limit} and \textbf{intrinsic complexity} to the software development practice.

\begin{itemize}
    \item \textbf{Limit.} To cope with software complexity, CDD requires developers to flag each point of complexity inside a code unit. For example, suppose a code unit's number of points of complexities (inspired by Magical Number Seven) is greater than the stipulated limit. In that case, developers should refactor the code and move the complexity to other code units. 
    
    \item \textbf{Intrinsic complexity.} CDD recognizes that the intrinsic load in source code (inspired by CLT) impacts developers differently. For a moment, suppose we have two developers working on the same code unit. Although they could likely disagree on what code elements hinder their understanding, this team of developers should curate a list with a few information items in the source code that they concur could make the code more complex.
\end{itemize}

By the disciplined use of CDD, we believe that developers could design smaller and easier-to-test code units.

\subsection{CDD Practice}\label{sec:cdd:practice}

Developers are frequently impacted by cognitive overload during the software development process~\cite{hermans2021programmer}. Usually, there are too many items of complexity that developers cannot efficiently process. More concretely speaking, in terms of source code, the programming constructs (their relationships) can be seen as our intrinsic items of complexity. Therefore, the first step in using CDD in a software development project is to define the Intrinsic Complexity Points (ICPs).

ICPs are code elements that could affect developers' understanding according to their usage frequency. A few examples of items of complexity include: code branches (e.g., \texttt{if-else}, \texttt{for} loops, \texttt{switch-case}, \texttt{do-while}, etc), conditionals, exception handlers (e.g., \texttt{try-catch-finally}), high order functions, inheritance, etc. These are items of complexity because, when used (or abused), they may burden developers' cognitive load.

The ICPs are though not limited to the ones presented above.
Indeed, the beauty of CDD is that developers can list any code element that could hinder their understanding, including testing code, SQL instructions, Infrastructure as a Code instructions, and the like (more on this in Section~\ref{sec:limitations}). For the sake of this work, we focused on Java-based ICPs.

After the selection of the ICPs that make sense to the development team, the team must define the cost of each ICP. Based on the Magical Number Theory, we believe that the overall cost of the code unit should be small. Therefore, our experience suggests that the individual cost of an ICP could vary from 0.5 up to 1, while the maximum cost of a code unit should not be greater than 10. If so, it is a flag indicating it is time to refactor to reduce its complexity.



\subsection{CDD Characteristics}\label{sec:cdd:characteristics}

CDD has at least two essential characteristics, which we present next.

\begin{enumerate}
    \item \textbf{CDD is straightforwardly measured.} While other design practices are described as abstractions and metaphors, CDD relies solely on counting the number of ICPs. Therefore, it is not only easier to compute CDD-related bad smells, but they are also easier to communicate. 
    
    \item \textbf{CDD is flexible:} Philosophically speaking, CDD could be seen as a design theory rather than design practice. This is because CDD gives to practitioners a general guide on what they may compute within a given code unit. Thus, as any general guide, the decision of what to compute is left to the practitioner (and her team). The only hard constraint CDD places is regarding the disciplined use of the limit. Any code unit that is over the limit must be refactored.
\end{enumerate}

\section{CDD at Zup}\label{sec:apporach}

In this section we describe our context at Zup Innovation (\S~\ref{sec:context}) and our approach to use CDD to build \handora (\S~\ref{sec:handora}). 

\subsection{Context}\label{sec:context}

Zup Innovation\footnote{https://www.zup.com.br/} is a sizeable Brazilian tech company. It currently has $\sim$3.5k employees, which work distributed mainly over Brazil, but also from other countries (from North America and Europe).
As a way to attract novice developers, Zup has maintained several bootcamps --- that is, short-term accelerating programs --- focusing on training soft and hard skills to novice developers. To join the bootcamp program, these novice developers participate in a fierce selection process. The selected participants were hired and become immediately part of Zup staff. Since they are hired, they are expected to dedicate full time to the training process.
These new hires spent 3--4 months in the training program, and during this period, no production-ready code is written; novice developers are expected to focus on their learning process. 
After the training program, the novice developers join real software development teams in the company.
In 2021 alone, 12 bootcamps were delivered, and more than 300 novice developers were hired through these bootcamp programs. 

\subsection{Building \handora}\label{sec:handora}

To support our tailored training process, we built our learning management system, called \handora\footnote{https://handora.zup.com.br/}. Unlike other online training tools, \handora was designed to handle the bootcamps' specificities, such as the flexible duration and the synchronous and asynchronous events. \handora is architectured as a set of six services\footnote{Given their granularity level, we cannot consider them as microservices.}. A general overview of \handora is present in Table~\ref{tab:handora}.

\begin{table}[h]
\caption{Overview about \handora's services. Column ``SLOC'' includes testing code while column ``Commits'' excludes merge commits.}
\label{tab:handora}
  \centering
  \begin{tabular}{lccrrr}
    \toprule
    \# & Lang. & Description & SLOC & Code Units & Commits\\
    \midrule
    S1 & Java & Core service & 16K & 230 & 767 \\
    S2 & Java & Search service & 3K & 36 & 115 \\
    S3 & Java & API provider service & 12K & 183 & 103 \\
    S4 & Java & Grading service & 1K & 40  & 66\\ 
    S5 & TypeScript & Frontend service & 18K & 79 & 334 \\ 
    S6 & Python & ML grading service & $<$ 1K & $<$ 10 & $<$ 10\\ 
    \bottomrule
  \end{tabular}
\end{table}


Four of the six services are written in Java, which are explored in this research. The TypeScript one (the frontend service) and the Python one (a machine learning service that provides a complementary grading process for the students' assignments) were not developed using CDD, so they are not the subject of our investigation. 
We discuss this limitation at Section~\ref{sec:limitations}. Moreover, the software development team used CDD only to design and implement the production code; the testing code was not written using the CDD principles. This is also part of our limitations.

\subsubsection{\handora's Team}

The software development team comprises eight engineers: one CTO, one senior engineer, one mid-level engineer, one mid-level UX designer, one senior UX designer, one UX researcher, one mid-level data scientist, and one senior infrastructure engineer. The team work in a distributed, remote-first environment. Asynchronous communication is often preferred over synchronous ones, although the team meets synchronously for $\sim$1 hour almost every weekday.

\subsection{CDD in Practice}\label{sec:cdd-at-zup}

In this section we discuss how we used the CDD principles to guide the development of \handora. 

\subsubsection{Defining the ICPs}\label{sec:cdd:icps}

The first step towards using CDD is defining the ICPs. The ICPs can be seen as language constructs that, if abused, could hinder one's understanding. The \handora development team decided to compute five kinds of ICPs. Since the team had to annotate the ICPs manually, choosing five ICPs would provide enough ICPs to use the practice, while not burdeding the developers with the manual effort. The selected ICPs are: 

\begin{enumerate}
    \item \textbf{Code branches:} this includes \texttt{if-else},  \texttt{switch-case}, the use of ternary operator, and all kinds of loops. This ICP resembles the cyclomatic complexity metric. The team counts 1 point for every branch. That is, while one \texttt{if} counts 1, one \texttt{if-else} counts 2.
    
    \item \textbf{Conditions:} this computes the number of condition in \texttt{if}, loops, etc. The team believes that conjoined conditions in an \texttt{if} statement could potentially hinder code comprehension. The team counts 1 point for every condition. For instance, the expression \texttt{if (a > b \&\& c < d)} computes 3: 1 for the \texttt{if}, one for the Boolean expression \texttt{a > b} and 1 for the other Boolean expression \texttt{c < d}.
    
    \item \textbf{Exception handling:} this computes the use of \texttt{try-catch-finally} blocks. The team counts 1 point per block. For instance, a code snippet that contains a \texttt{try-catch-finally} blocks would count 3 ICPs: 1 for the \texttt{try} block, 1 for the \texttt{catch} block, and another one for the \texttt{finally} block.
    
    \item \textbf{Internal coupling:} this includes the uses of classes of the same project. We compute ICPs only for a subset of the domain classes, for instance, classes that deal with the database are considered. The team counts 1 point when using these classes.
    \item \textbf{External coupling:} this includes the use of code units from the JDK or external libraries and frameworks. Like internal coupling, the team considered only a subset of these external classes. Unlike internal coupling, though, the team counted only variable declarations. The team counts 0.5 points for each external coupling.
    
\end{enumerate}

Regarding the costs, except for \emph{external coupling}, all other ICPs had the same cost, that is, there was no weighting in the ICPs.
Still, other teams could consider other set of ICPS. For instance, there was a long discussion about whether the team should consider using lambda expressions as part of the ICPs. While one team member argued that using lambda expressions could hinder his understanding, two other engineers had different opinions. Therefore, they decided not to consider lambda expressions as an ICP. In summary, the team should find consensus regarding the set, the cost, and the limit of the ICPs. If engineers disagree or were neutral, the CTO often gave the final answer.

\subsubsection{Defining the costs of each ICP}

Afterward, one should also define the cost of each ICP. As a general rule of thumb, each ICP costs 1. However, the team could vary this definition accordingly. After a long discussion, the \handora team decided that external coupling cost 0.5 only. The rationale is that even though the team frequently deals with these cases, they still spend a few seconds trying to figure out their meaning; all other ICPs cost 1.

\subsubsection{Defining the limit} 

The limit is an essential concept in CDD. It does not only indicate that one given class has more ICPs than expected, but it also requires developers to refactor that class to reduce its complexities. 
The development team was taught that a class over the limit conveys the same meaning as a class that does not compile: something is not right and should be fixed. 
The software development team decided 10 as the upper bound limit. This limit was a first suggestion that the team adopted, but did not change throughout the software evolution.
Therefore, any class with more than 10 ICPs must be refactored; either to remove the ICPs or to extract the complexity of the ICPs to another class. 


\subsubsection{Computing the ICPs} Since the team did not employ tools for searching and computing the defined ICPs in the source code, the team had to compute them manually. To do so, the engineers created the \texttt{@ICP} annotation, which they used to annotate each ICP instance within a class. This facilitated the process of counting the ICPs because the team used the search features of the IDE. After computing the individual ICPs, the class is annotated with the total ICPs. An example of how the team calculated the ICPs is presented in Listing~\ref{lst:icps}.

\begin{figure}[htbp]
\begin{center}
\begin{minipage}{0.8\textwidth}
\begin{lstlisting}[escapechar=!]
@RestController
@RequestMapping("/certificates")
@ICP(8)
public class CertificateDetailsController {

  @ICP(1)
  private CertificateRepository repo;
  @ICP(1)
  private TrainingCompleted trainingCompleted;

  @ICP(2)
  @GetMapping("/{certificateId}")
  public CertificateResponse execute(
                    Long id, Student student) {
    @ICP(1)
    var potentialCertificate = repo.findById(id);
    var certificate = potentialCertificate.get();
    
    @ICP(2)
    if (certificate.doesNotBelongTo(student)) {
      throw new ResponseStatusException(NOTFOUND);
    }
    @ICP(1)
    var training = certificate.getTraining();

    return trainingCompleted.check(
        training, student, 
        () -> new CertificateResponse(certificate));
  }
}
\end{lstlisting}
\end{minipage}
\caption{A simplified version of an annotated class using CDD.}
\label{lst:icps}
\end{center}
\end{figure}

From top to bottom: the first \texttt{@ICP(8)} means that this class used 8 ICPs overall. Inside the class, the team flags every use of the selected ICPs. There are two \texttt{@ICP(1)}, one for each variable: \texttt{CertificateRepository repo} and \texttt{TrainingCompleted trainingCompleted} (indicating an internal coupling). 
On top of the \texttt{execute} method, there is \texttt{@ICP(2)}, indicating an internal coupling with the \texttt{Student} class (a parameter) and another one with the \texttt{CertificateRespone} (the response type); the team does not consider the use of the \texttt{Long} type as a form of coupling. 
The next use of \texttt{@ICP(1)} is regarding the coupling with the \texttt{repo} var (from the \texttt{CertificateRepository} class). 
There is also one \texttt{@ICP(2)} for using an \texttt{if} statement (1 ICP) and for having one condition (other 1 ICP). 
The final \texttt{@ICP(1)} refers to the coupling with \texttt{training}. Note that the lambda expression was not annotated because the team decided not to consider it an ICP.


\subsubsection{Revisiting the ICPs} The development cycle could be summarized as ``1+4 weeks'': the first week is dedicated to planning activities while the other four weeks are for building activities.
The first week is also known as the ``slow down'' week, where the engineers could relax a bit from the four building weeks. During this planning week, the engineers also had one regular spot to discuss the use of CDD. During this meeting, the engineers shared the decisions they made to keep the classes within the limit, and the challenges in doing so. The meetings were recorded and analyzed in this study.
During this study, we observed $\sim$40 cycles

\section{Research Methodology}\label{sec:data}

The study is primarbased on an observation of the software development team: while the first author was not a participant of the team, the second author played the CTO role in the team. We complement these observations with data and metadata from the repositories, data from the team's meetings, data from the online documents, and others.  

\subsection{Data Collection}
The data collected in this work covers October 2021 (when the software development project started) until September 2022, when the first release of the product was made. We collected data from three different sources: git logs, ADRs documents, and video retrospectives. 

\subsubsection{Mining \texttt{git} logs}

We mined \texttt{git} logs in order 1) to understand how classes evolved over time, and 2) to find and understand CDD-driven code changes. 
More concretely, we analyzed all commits performed in the \texttt{main} branch of the three studied services, totaling 985 commits; we excluded merge commits. These commits were performed by a team of five developers. For each commit, we then performed \texttt{git checkout} to restore the working tree files, and then we run a set of tools, such as Spoon~\cite{pawlak:SPE:2015}, to compute general coding and CDD metrics, and Refactoring miner~\cite{Tsantalis:TSE:2020}, to assess the presence of refactorings.
As we shall see in Section~\ref{sec:findings}, this mining part contribute to understanding how CDD was used, and how it impacted \handora's development.

\subsubsection{Revisiting the ADRs}

We revisited the Architecture Decision Records (ADRs), which are text-based documents describing the main design/architectural decisions made during the software development process. ADRs are versioned in git, but in a different repository.
The first authors revised these documents and employed a qualitative coding technique to categorize them. On average, one ADR has 138 words. Among the 8 ADRs, just one contained CDD-related information. In this particular ADR, the development team updated the set of ICPs used, and the decision to use them.

\subsubsection{Rewatching the CDD retrospectives}
We then watched the recorded online meetings in which the team reflected on the use and adoption of CDD. Although we had a regular spot for this meeting during the planing week, only six videos were found and analyzed.

\subsubsection{Data Analysis}

The data analysis was performed as follows: the first author studied the documents found and designed initial observations. The second author, who was part of the software development team, could confirm or refute those observations. This author eventually brought up his perception, which was not always documented. Whenever we have potential pieces of evidence about the benefits/challenges in the use of CDD, we present them to the remaining engineers in the team. 
To mitigate potential biases, we first asked their perception (e.g., ``do you think CDD could also impact the size of the testing methods?''), and only after their response we showed our findings. We also asked the engineers individually, in no particular order. 
Three engineers gave regular feedback, which we used to refine our claims. 
Before submission, this report was read by the whole software development team. Since the software engineering team was unaware of the conclusions reached in this study, we could validate the findings we drew from the data during this process. In general, the engineers agreed with our observations.

\section{Findings}\label{sec:findings}

In this section we group the main findings of this study.

\subsection{CDD limits the size of code units}

\begin{figure}
  \centering
  \includegraphics[width=\linewidth, clip, trim= 0px 40px 0px 50px]{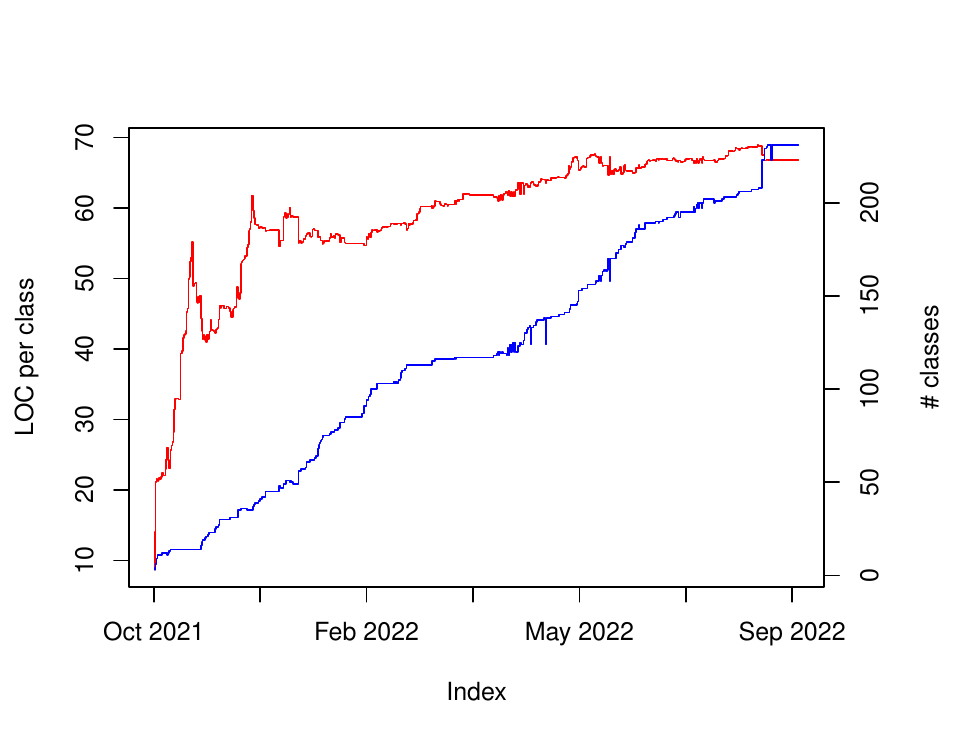}
  \caption{The average of 1) the sizes (in terms of SLOC) of the classes (\colorbox{red!25}{red} line) and 2) the number of created classes throughout software evolution (\colorbox{blue!25}{blue} line). We used the \texttt{wc} UNIX tool to calculate lines of code. We considered blank lines and comments. We did not consider testing classes here.}
  \label{fig:handora}
\end{figure}

In Figure~\ref{fig:handora}, the \colorbox{red!25}{red} line shows the evolution of the averaged lines of code. In contrast, the \colorbox{blue!25}{blue} line shows the evolution of the number of classes created during the evolution of the project.

Here we noticed our first striking observation: the number of lines of code (\colorbox{red!25}{red} line) per class dramatically increases at the beginning of the project. Then, however, around 1/5 of its evolution, the growth starts to plateau. 
On the other hand, we could notice a nearly linear growth of the number of classes (\colorbox{blue!25}{blue} line). Taken together, these two lines indicate that code units can be kept under the limit, even with the (near) linear growth of the software product. We presented this finding to the software development team, who concurred that this was primarily due to the CDD practice. 


\begin{wrapfigure}[13]{r}{0.3\linewidth}
    \centering
    \vspace{-0.2cm}
    \includegraphics[width=.7\linewidth, clip, trim= 30px 70px 30px 50px]{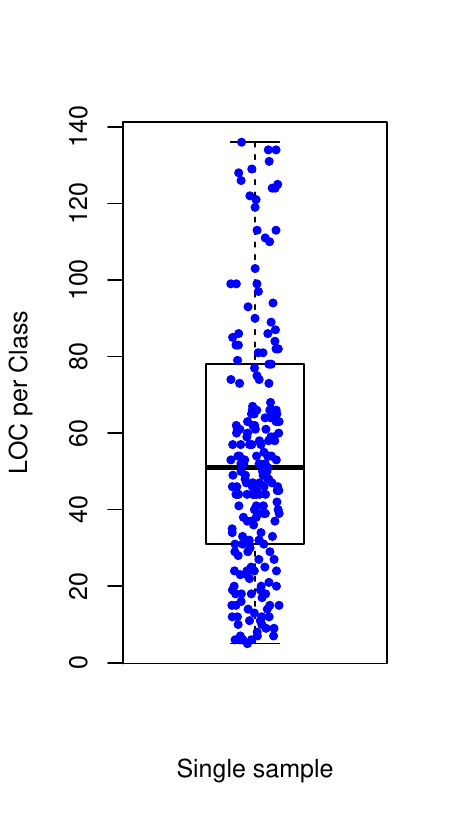}
    \caption{\label{fig:loc} SLOC distribution in the latest \handora release.}    
\end{wrapfigure}
We then took a closer look at the latest \handora release. Figure~\ref{fig:loc} shows the result.
We noticed that, on average, an \handora class has 59 lines of code (17.2 standard deviation).
This observation corroborates with our previous findings, indicating that CDD could impact the size of the classes. 
The software development team reinforced that the small size of the classes is probably due to the limit that CDD places. One of the team members also indicated that if the limit was, say, 30, the classes would probably be more extensive. This suggests that the disciplined use of CDD could drive the lengths of the classes.


\MyBox{\textbf{Key observation \#1:} By using CDD, the software team could keep code units under the limit, even with the (near) linear growth of the software.}

\subsection{CDD's limit is not written in stone}

\begin{figure}[h]
  \centering
  \includegraphics[width=\linewidth, clip, trim= 0px 40px 0px 50px]{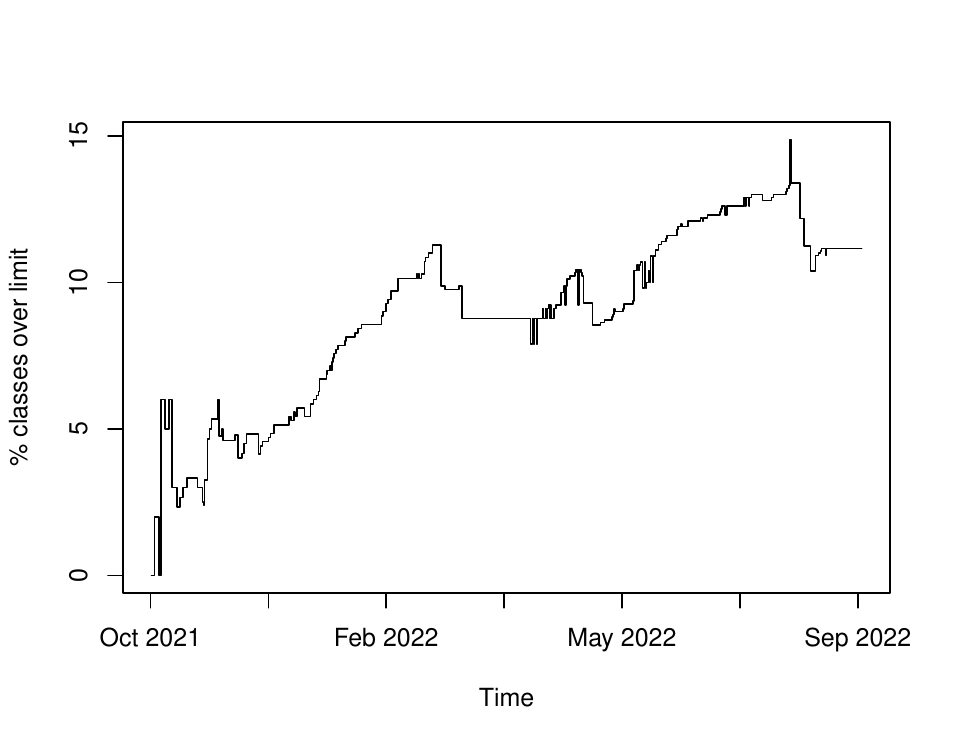}
  \caption{Percentage of classes over the limit. We did not consider testing classes here.}
  \label{fig:classes-over-limit}
\end{figure}

There are, obviously, classes that are bigger than the stipulated limit. Figure~\ref{fig:classes-over-limit} shows the percentage of classes over the limit during the software evolution.

As shown in this figure, the percentage of classes over the limit increases over time. At the peak, there were 14\% of classes over the limit. Although the software development team strongly advocated in favor of CDD, we observed some scenarios in which the development could not reduce the complexity of the class to fit into the limit. Two examples are following:

\begin{itemize}
    \item \textbf{Core classes:} \handora is an online learning management system, and as such, it is based on several learning theories. The classes that implement these learning theories are also less likely to stay under the limit because they are core classes in the system, and as such, they have to cope with many responsibilities.
    \item \textbf{Classes with rich contracts:} \handora is built as a set of orchestrated services. These services talk to each other through well-established contracts, implemented as Data Transfer Objects (DTOs). Though contracts cannot be reduced or simplified, and thus the DTO classes are less likely to stay under the CDD limit. 
\end{itemize}


We discussed these cases with the development team. To minimize the size of \textbf{core classes}, the team adopted the use of \emph{partial classes}.
A partial class is a feature available in C\#, which the team incorporated into the product. By using partial classes, a single class's functionality can be split into multiple files, and these files are merged into a single one.
However, the use of partial classes also brings shortcomings. For instance, when splitting a class into three files, the team used the following naming convention: \texttt{PartialClass\$1.java}, \texttt{PartialClass\$2.java}, and \texttt{PartialClass\$3.java}. Although the team was able to come up with better names in some cases, the above naming convention was still common.
The development team acknowledged that this convention could potentially hinder the understanding of a new developer onboarding the team. Moreover, these classes are grouped in isolated packages; therefore, navigating between these classes might not also be straightforward.


The team also reported that they could not use the partial class approach to refactor the \textbf{classes with rich contracts}. This happened because the contracts are in JSON format, and the JSON document is automatically built using the public attributes defined in the classes. If the team split the classes into several files, this would generate JSON documents with different formats, impacting the clients. Knowing that DTO classes would hardly fit in the 10 ICPs limit, the development team decided that, for these kinds of classes, the limit would be 20 ICPs.

\MyBox{\textbf{Key observation \#2:} The team acknowledged that keeping code units under the limit is not always feasible, but new strategies were proposed to deal with these cases.}

\subsection{CDD impacts ICP size}

\begin{figure}[h]
  \centering
  \includegraphics[width=\linewidth, clip, trim= 0px 40px 0px 50px]{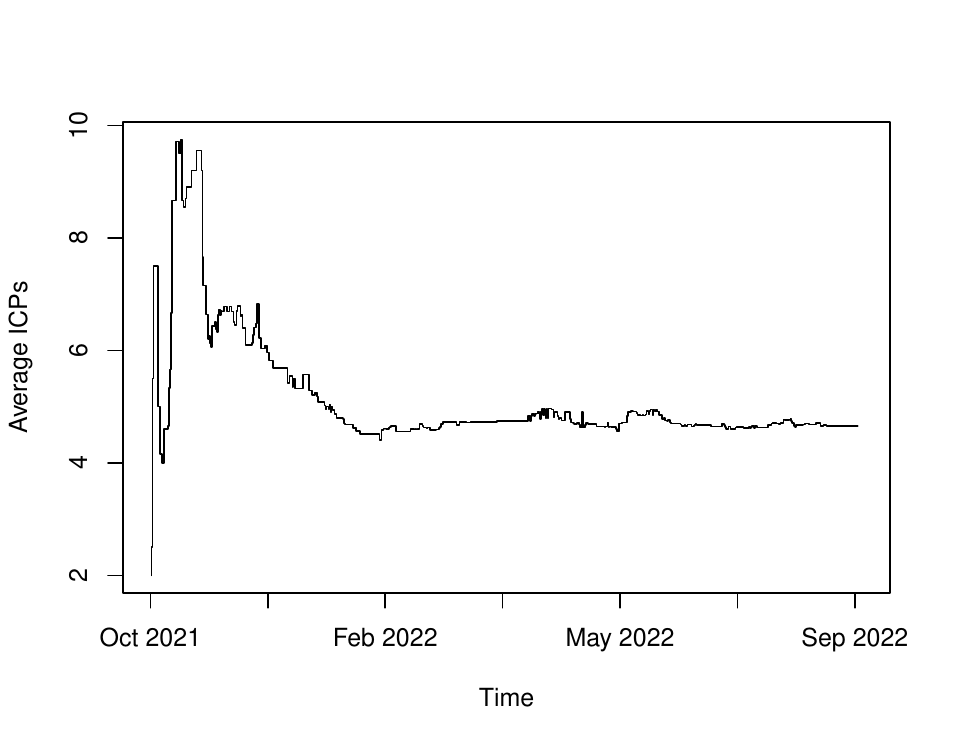}
  \caption{The averaged value of ICP per commit. We did not consider testing classes here.}
  \label{fig:icp}
\end{figure}

Another way to observe the growth of complexity over time is by computing the number of ICPs per code unit. In Section~\ref{sec:cdd:icps} we described the ICPs considered by the software development team. To compute these ICPs, we built a static analysis tool
using Spoon~\cite{pawlak:SPE:2015}. Figure~\ref{fig:icp} shows the average number of ICPs per class throughout software evolution.

This figure shows a similar behavior observed in Figure~\ref{fig:handora}: it seems that in the very first commits of \handora, there is a high fluctuation in the use of computed ICPs. However, as time passes by, there is also a sort of plateau. 
There seems to be a correlation between the size of classes and the number of used ICPs. 

\MyBox{\textbf{Key observation \#3:} There seems to be a correlation between the size of the class and the number of ICPs used in the class.}

\subsection{CDD impacts method size}

According to Gill and Kemerer~\cite{Gill:TSE:1911}, the module size can indicate code quality since maintenance effort is positively correlated with method length.
Moreover, Chowdhury and colleagues~\cite{Chowdhury:MSR:2022} discovered that methods with 24 lines of code or fewer are less prone to bugs for open-source Java projects. In conclusion, the authors suggested that ``developers should strive to keep their methods within 24 SLOC''. We used this evidence to shed some light on whether CDD might also influence the quality of the size of the methods in \handora.

\begin{figure}[!h]%
 \centering
 \subfloat[S1]{
 \includegraphics[width=.35\textwidth, clip, trim= 0px 0px 30px 80px]{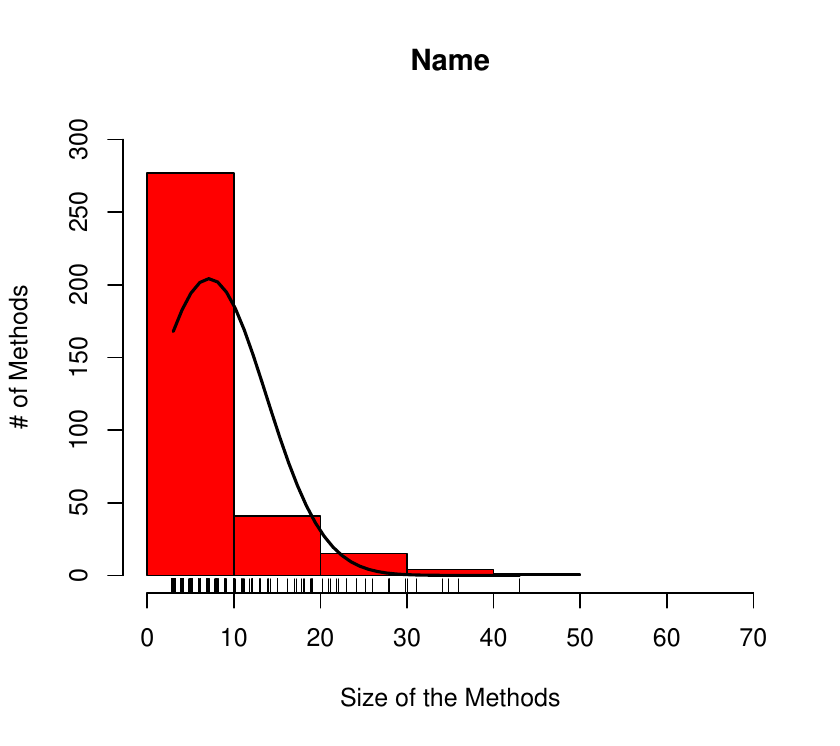}
 }%
 \subfloat[S2]{
 \includegraphics[width=.35\textwidth, clip, trim= 0px 0px 40px 80px]{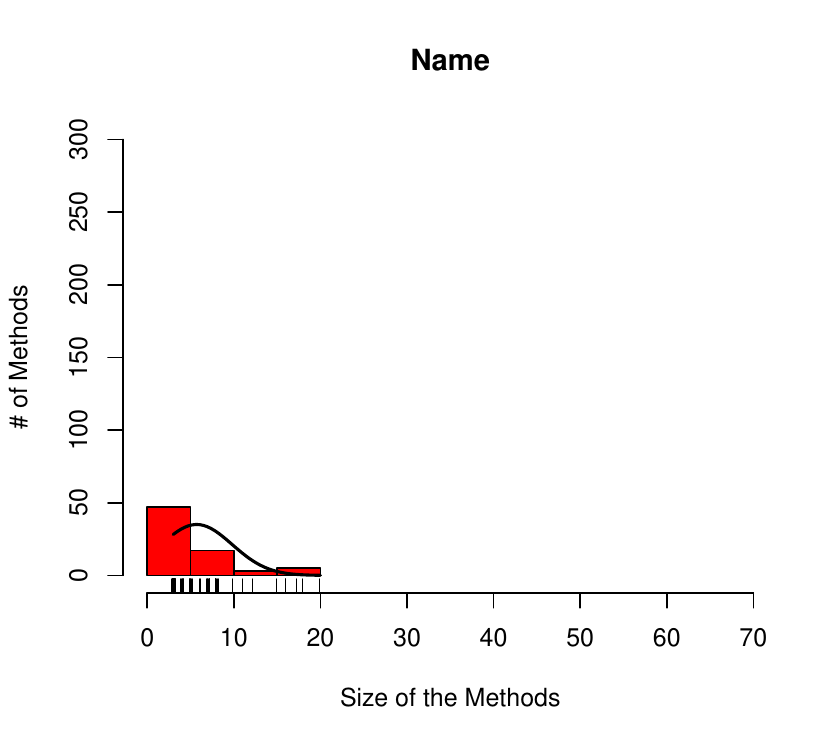}}
 
  \subfloat[S3]{
 \includegraphics[width=.35\textwidth, clip, trim= 0px 0px 30px 80px]{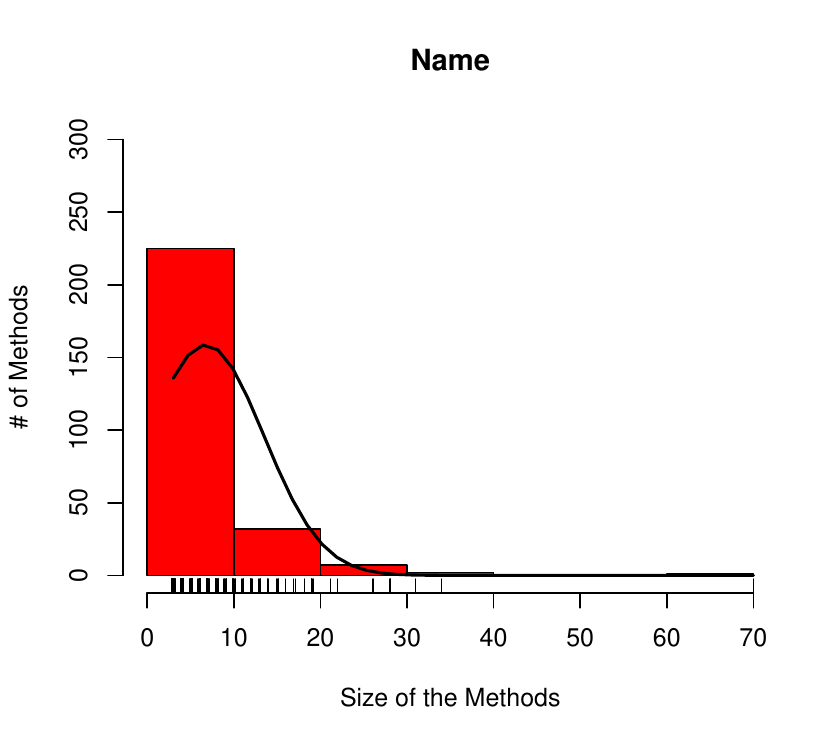}}
 
  \caption{Distribution of the methods' lengths at \handora. This figure does not include testing methods.}
 \label{fig:msize}
\end{figure}

To do so, we extracted data from S1, S2, and S3 only because they are the services with the greatest number of code units and methods. Following mining best practices, we filter out \texttt{getters} and \texttt{setters} methods, testing methods, as well as \texttt{hascode()} and \texttt{equals()} implementations, because they could add noise to the data~\cite{Nahla:ICSE:2019}. Overall we removed 525 methods; 677 methods remained for analysis. In particular, S1 had 230 code units and 338 methods, S2 had 36 code units and 72 methods, and S3 had 183 code units and 267 methods.
Figure~\ref{fig:msize} depicts the distributions found.

As one can observe, overall, the majority of the methods in these services are under the 24 SLOC limit. More concretely, for S1, 96.4\% of the methods are under the 24 SLOC limit (S2: 100.0\%; S3: 97\%). On average, a method at \handora has 6.8 lines of code (median: 4; standard deviation: 6.4). 
We then asked the development team whether they see any influence on the use of CDD on the size of the methods. They concur that the size of the methods might also be influenced; one of the engineers commented that: ``Every unit of code is impacted, because we know what the limit is and what goes into that limit.''.
The most extensive method at \handora has 70 lines of code, at S3. This particular method implements a visitor pattern that navigates through Markdown files. This method computed 8 ICPs, whereas the class computed 10 ICPs overall. We also asked the development team whether they had any intention of refactoring this particular method. The engineers said that since the class is under the limit, there is no purpose in refactoring a method, even if a large one. Another engineer noted that the use of CDD highly influences the decision of when to refactor. Therefore, they hardly refactor classes/methods that are under the limit.

\MyBox{\textbf{Key observation \#4:} By using CDD, achieving the 24 lines of code threshold becomes a reasonable easy task for the software development team.}

\subsection{CDD impacts testing code}\label{sec:testing}

As mentioned in Section~\ref{sec:handora}, 
although the team did not use CDD for testing purposes, here, we also aimed to understand whether CDD also impacted the testing code.
Table~\ref{tab:testing} shows a summary of the testing code at \handora. 

\begin{table}[tb]
\caption{Overview of testing characteristics at \handora. The column ``Testing Units \%'' means the percentage of testing classes compared to the overall classes at \handora. Coverage data was calculated using JaCoCo.}
\label{tab:testing}
  \centering
  \begin{tabular}{lcccccc}
    \toprule
       &      & \multicolumn{2}{c}{Testing Units} &  & \multicolumn{2}{c}{Coverage} \\
    \cline{3-4}\cline{6-7}
    \# & SLOC & \# & \% &  \# Methods & Instr. & Branch  \\
    \midrule
    S1 & 7.6k & 77 & 33.4\% & 215 & 66\% & 71\% \\
    S2 & 1.3k & 19 & 52.7\% & 41  & 37\% & 61\% \\
    S3 & 5.2k & 53 & 28.9\% & 128 & 58\% & 64\% \\
    S4 & 0.3k & 7  & 17.5\% & 25  &  2\% &  0\% \\
    \bottomrule
  \end{tabular}
\end{table}


As shown in the table, except for S4 (the minor service), all other \handora's services have reasonably good code coverage. By means of comparison, code coverage at Google is around $\sim$80\%~\cite{ivankovic2019code}. On the other hand, the proportion of testing code units is reasonably smaller than the production code units. On average, there are 3.2 testing methods per testing unit, while there are 2.4 methods per production code unit. That is, \handora has fewer testing units, although these testing units tend to concentrate a higher number of methods, when compared to production code units.
In terms of size, on average, testing methods have 8.1 lines of code (median 7; max 35; standard deviation: 5), which is inline with recent research that suggests that small testing methods (10 SLOC or less) as a good testing practice~\cite{Kochhar:SEIP:2019}. 

Table~\ref{tab:testing-comparison} compares the production code's size and the testing code's size. In this table, we can observe that, on average, the size of a testing method is generally more significant than the production method (8.1 SLOC vs 6.8 SLOC, respectively). On the other hand, production methods have a longer tail, meaning that production code units have larger methods. This finding might also be aligned with the study of Kochhar and colleagues, which observed practitioners concurred that ``a good suite contains lots of small test cases (with fewer LOC) and few large test cases''~\cite{Kochhar:SEIP:2019}.

\begin{table}[h]
\caption{Descriptive statistics on the number of all production methods and testing methods at \handora.}
\label{tab:testing-comparison}
  \centering
  \begin{tabular}{lcccccc}
    \toprule
        &    Min & Avg & Median & Max & SD & Histogram\\
    \midrule
    Production  & 3 & 6.8 & 4 & 70 & 6.4 & \includegraphics[scale = 0.1, clip = true, trim= 62px 75px 0px 50px]{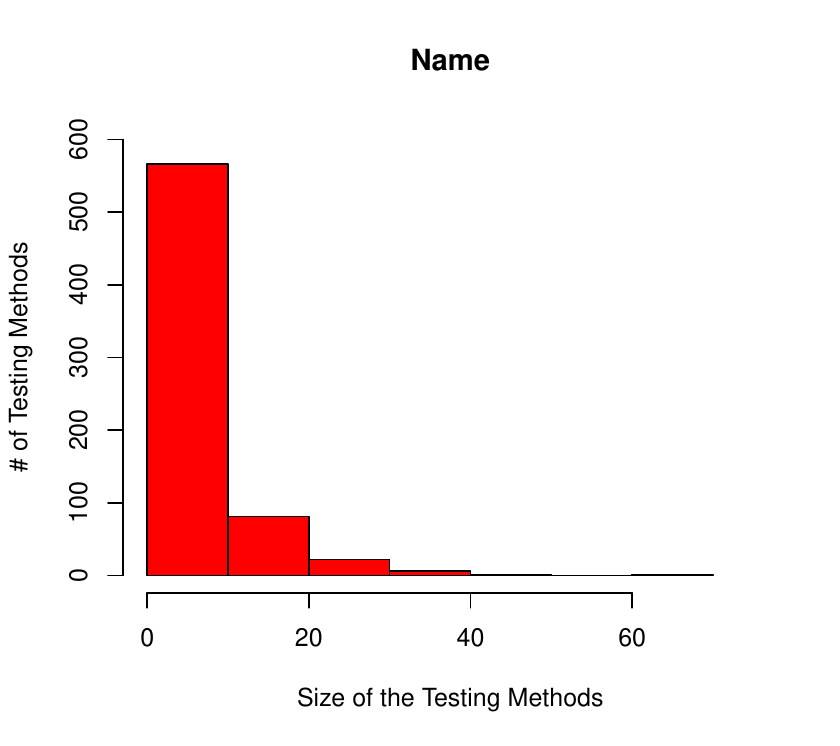}\\
    Testing     & 3 & 8.1 & 7 & 34 & 5.0 & \includegraphics[scale = 0.1, clip = true, trim= 62px 75px 0px 50px]{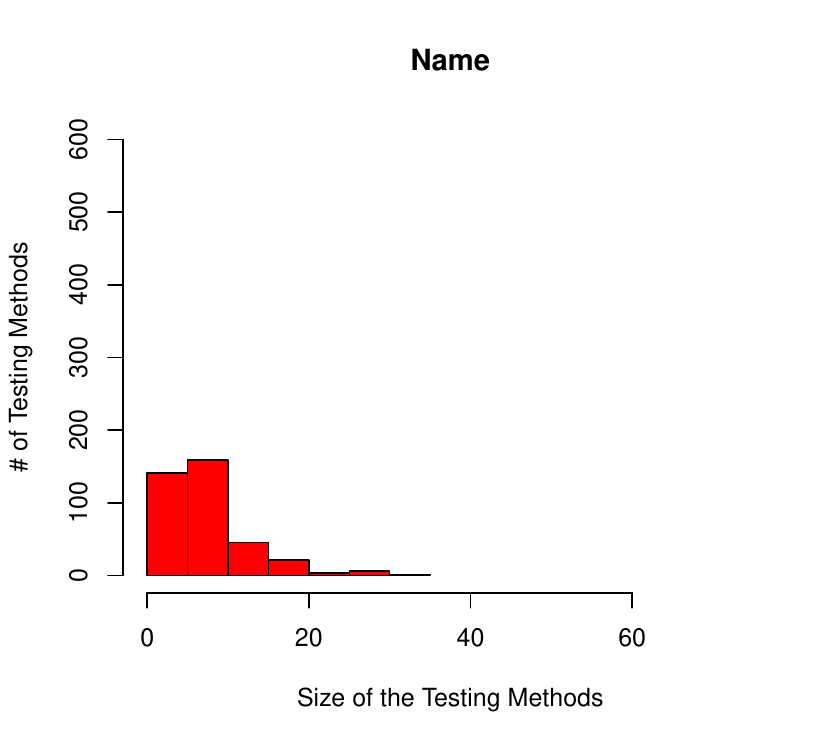}\\
    \bottomrule
  \end{tabular}
\end{table}


We also asked the team about the testing effort. One of the engineers mentioned that creating new testing code---at the latest release of \handora--- somehow requires \emph{less} effort than the effort they placed at the project's beginning. The engineer complemented by suggesting that this might be due to the reuse of the existing testing infrastructure. When we asked whether the size of the production classes might influence their testability, one engineer said, ``I think there is a relationship because the complexity of the test can be seen as a proxy of the complexity of the code under test''. Another engineer complemented that ``there is an indirect relationship between CDD and the size of the testing code. Since CDD leads to small code units, these code units are also easier to test.''

One of the engineers also complemented that ``penalizing conditionals tends to ease the test writing process''. This happens because, according to more sophisticated coverage criteria such as MC/DC~\cite{yu2006comparison}, the higher the number of conditions, the higher the number of test cases required to cover them. We found that, on average, there are 1.3 conditions per \texttt{if} statement in the codebase.

Finally, we tried to assess any association between the size of the production methods and the testing methods.  
Since these sets of production and testing methods have different lengths, we \emph{downsampled} the production set to match the size of the testing set. 
We ran the Shapiro-Wilk, which indicated that both sets are not normally distributed (production = 0.63494, p-value = 0.0001; testing = 0.80003, p-value=0.0001). We ran the Person correlation because it does not assume normality. Person correlation, though, indicated no association exists between the size of the two sets of methods (p = -0.09742104).


\MyBox{\textbf{Key observation \#5: } Testing code is consistently smaller than production code. The team recognized that the effort to create new testing code is reasonably small, which might be due to the size of the production code.}

\subsection{CDD-driven commits}\label{sec:cdd-commits}

During the software development process, the team was also instructed to perform commits with the pattern ``\texttt{cdd(class): description}'' whenever they needed to perform changes to adhere to the CDD principles. We mined these commits to shed additional light on the developers' intentions when performing them. We found 38 commits with this directive: 32 in S1 and 6 in S3. We found no CDD commit in the other services. While manually examining these commits, we discovered a few intentions:

\begin{itemize}
    \item \textbf{Annotating classes}. The development team does not always use the \texttt{@ICP} annotation. In the beginning of the project, the value of the ICP was documented as a comment in the source code. After introducing the \texttt{@ICP} annotation, the team had to annotate the unannotated classes. 
    \item \textbf{Recomputing ICPs}. Whenever a new point of complexity is added to the code, the team has to recompute the class's ICPs. This kind of commit usually performs minor changes in the source code, in addition to the new ICP value for the class. 
    \item \textbf{Committing feature first, CDD later}. The software development team was also instructed to focus first on completing the coding task. After implementing a new feature, the team refactored the code to adhere to CDD. To avoid technical debt, the team was also instructed not to delay these refactorings. The team intended to make the code units fit under the stipulated limit in these kinds of commits. 
    
\end{itemize}

Considering that the team often performs these CDD-driven commits as a way to conform the code to the CDD guidelines,
we hypothesized that these commits are also more likely to perform refactoring operations.
We then ran RefactoringMiner~\cite{Tsantalis:TSE:2020}, the state-of-the-art tool in finding refactoring operations in commits. 
RefactoringMiner found 29 kinds of refactorings, performed 266 times at \handora. On average, one CDD-driven commit includes seven refactoring operations.
Among the 29 kinds of refactorings, 11 (38\%) are annotation-based refactorings, mostly due to the need to adapt the \texttt{@ICP} annotation. The annotation-based refactorings include: Add Attribute Annotation (25 instances; 9.3\%), Modify Class Annotation (26 instances; 9.7\%), and Remove Parameter Annotation (17 instances; 6.4\%). 
Other than annotation-based refactorings, we also found a few instances of class-based refactorings (8.6\%), such as Move Class (7 instances; 2.6\%), Rename Class (12 instances; 4.5\%), and Extract Class (4 instances; 1.5\%). Indeed, these refactorings tend to be less common since they touch on more stable parts of the code (i.e., class declarations). We then compared the presence of these class-based refactorings in the other \handora commits. On average, an ordinary commit in \handora includes 2.6 refactoring operations. 
Class-based refactorings also appear less often: overall, they represent 6.8\% of the refactorings performed. The class-based refactorings found are: Change Class Access Modifier (12 instances; 0.6\%), Extract Class (12 instances;  0.6\%), Move Class (45 instances; 2.3\%), and Rename Class (63 instances; 3.2\%).

\MyBox{\textbf{Key observation \#6:} By instructing the team to finish the task first and then adhere to CDD practices, developers tend to perform refactoring operations more often.}





\section{Lessons learned}\label{sec:lessons}

In the previous section, we report the team's experience using CDD to build a real software product at Zup Innovation. In this section, we reflect on this experience and extract a set of lessons learned that could be useful for practitioners interested in adopting CDD. 

\vspace{0.1cm}
\noindent
\textbf{CDD flexibility is good.} As we presented in Section~\ref{sec:cdd:characteristics}, CDD has two unique characteristics: 1) it is flexible so that developers can choose the ICPs they believe are the most important ones. After defining, 2) the ICPs can be objectively measured. This work also highlighted only one hard constraint: the limit. However, we also noted that the software development team was able to flex it a bit. For instance, to cope with the rich contracts of the services, the development team decided to change the CDD limit of these kinds of classes (from 10 to 20). Based on this experience, we envision that software development teams could define different limit levels for different domain classes. 

\vspace{0.1cm}
\noindent
\textbf{CDD flexibility is bad.} Since CDD allows the team to decide the types, the costs, and the limit of the ICPs, these decision could become personal; other teams could decide differently. Although we presented the configuration adopted by the \handora's team, it is hard to provide a general guidelines.

\vspace{0.1cm}
\noindent
\textbf{CDD may increase developers' effort.} Although one of the main benefits of CDD is to \emph{reduce} developers' cognitive load, we also perceived that the engineering team had to place some effort to get used to CDD. One of the engineers also said that, besides being using CDD for over a year, CDD is not yet straightforward to him. Moreover, by mining CDD-driven commits (Section~\ref{sec:cdd-commits}), we also noticed several commits with the intention of updating the \texttt{@ICP} annotation, without further changes. This indicates that developers may not always count the ICPs correctly and have to fix them.

\vspace{0.1cm}
\noindent
\textbf{The lack of tool support hampers CDD adoption.} The software development team at Zup had to compute the ICPs manually. The team reported that computing them is a tedious and (often) error-prone task, requiring a non-trivial mental effort. Although we have built a few CDD detectors, developers interested in adopting CDD still need production-ready tools that can be used in different programming environments. Not only for computing ICPs, but tools that could ease visualization, perform code transformations, or increase developers' awareness of refactoring opportunities; this could be impactful in contributing to CDD adoption.

\vspace{0.1cm}
\noindent
\textbf{CDD could better guide refactoring operations.} Even if developers are aware of the presence of code smells in the codebase, they may not always refactor the code to remove them~\cite{Silva:FSE:2016}.
By using CDD, we believe developers can perform refactoring operations more conscientiously, 
due to the notion of limit. The limit reinforces the need to refactor. Indeed, the software development team indicated that they only refactor a class to reduce its complexity \emph{if}
that class is above the limit; otherwise, they do not refactor. 

\vspace{0.1cm}
\noindent
\textbf{CDD helps in identifying potential god classes earlier.} God classes ``feature a high complexity, low inner-class cohesion, and heavy access to data of foreign classes. It violates the object-oriented design principle that each class should only have one responsibility''~\cite{olbrich2010all}. For developers, avoiding god classes is of paramount importance, and researchers have been studying approaches to detect and refactor them for quite some time (e.g.,~\cite{lanza2007object,marinescu2001detecting}). Our experience suggests that --- by design ---  CDD helps developers avoid god classes. This happens because CDD increases the awareness of the size of the code units so that any code unit getting big is clearly noticed in advance before it gets huge.

\vspace{0.1cm}
\noindent
\textbf{Software teams must exercise self-discipline.} Due to its manual characteristic, CDD requires a lot of discipline from the teams. As a consequence, teams that operate under a tight schedule might have a hard time trying to adopt this design practice. Indeed, the benefits that we perceived in using CDD encouraged us to advocate in favor of CDD to other software development teams at Zup. However, our experience so far in spreading the use of CDD was not successful. Even though the teams seemed to see the value in adopting CDD, these teams were operating behind schedule, so adopting a new design theory requires some time that these teams were unable to commit.

\vspace{0.1cm}
\noindent
\textbf{Limiting the size of a code unit through ICPs is better than through SLOCs.} One observant reader might argue that our approach could be even easier to employ if we consider the SLOC of a class as our notion of limit. However, our experience suggests that limiting the size in terms of SLOC could lead to biased results. This could happen due to at least two reasons: 1) small classes could still have a high level of complexity, and 2) SLOC may as well consider elements that do not add complexity to the code (e.g., comments, classes, and methods declarations, etc.). If developers have to filter out these elements, this could, in turn, hinder its adoption by developers.

\vspace{0.1cm}
\noindent
\textbf{There is no silver bullet set of ICPs.} Given the myriad of ICPs one could choose, the team had to work hard to define (and refine) the ICPs. 
The ones that the development team chose (Section~\ref{sec:cdd:icps}) were the ones that the team believed could ease maintenance efforts. As one engineer said: ``in my opinion, it is not about finding the perfect ICP, but finding the ones that work in that given context; and this is the beauty of CDD''. Therefore, engineers interested in adopting CDD could start from our set of ICPs, and adapt as needed.





\section{Limitations}\label{sec:limitations}


Unlike traditional MSR studies, which investigate an ever increasing number of data points, our study is based on observing a single team working on one single software product. \handora itself is not a mission-critical product. \handora's codebase is written in an established programming language, using popular frameworks. 
Although our findings contribute to understanding how CDD was helpful in creating \handora, we highlight that these findings would hardly generalize to other teams or products, either on Zup or from other companies.  It is also unclear how CDD could contribute to designing more complex software projects. 

Another limitation is regarding the team's experience using CDD to develop Java-based services. Although \handora has other services written in different programming languages (such as TypeScript and Python), the team decided not to use CDD in these additional services. The rationale for this choice is that, since the software development team was still getting acquainted with CDD, adopting it in different scenarios would burden the team more. Still, we did not checked whether the ICP annotations are complete (in terms of false negatives). Indeed, we noticed commits that solely changes the \texttt{@ICP} annotation, which indicates that developers may miscount the ICPs. 

Moreover, as discussed in Section~\ref{sec:cdd}, CDD flexibility allows one to compute any kind of ICP. In this paper, we limited our observations to Java-based ICPs. Still, we also understand that there are many other ICPs in other programming languages and frameworks that the team could consider. Still, we did not use CDD for designing testing code because the team was unsure which ICPs they should use and follow. It is unclear how the use of different ICPs could drive the software development process, in particular, when choosing ICPs that could be more complex to compute. We believe these limitations could drive new CDD-related research in the future. 

A final limitation is regarding how the team perceived the benefits of CDD. Since the team members are early adopters of the CDD practice, they might be positively biased towards CDD; for instance, being more motivated to produce high-quality code. 
Moreover, their interest in using CDD might have overcame the challenges. When we asked the team about the potential difficulties that CDD brings to the software development process, the team talked mostly about 1) the lack of tools and 2) the need to recompute the ICPs.
To some extent, this limitation was alleviated because these were professional programmers. Still, the team did not know they would be part of a research study.

\section{Related Work}


Software design practices are unquestionable tools in the programmer's arsenal. However, they also have known limitations. Take, for instance, three well-known metrics: 1) McCabe's cyclomatic complexity~\cite{mccabe1976complexity}, a metric that computes the number of independent paths of a program, Hasltead's complexity~\cite{halsted1977elements}, a vocabulary-based metric, and SLOC, that counts the source code lines of code. If we calculate these three metrics for the program shown in Figure~\ref{lst:icps}, these three metrics will yield different (perhaps conflicting) results. 
Since one of the goals of these metrics is to guide practitioners to parts of the code that need more attention, the different calculations these metrics perform could cause confusion. Moreover, in a systematic mapping study conducted by Varela~\cite{Varela:2017:JSS}, it was observed that only during the period from 2010 to 2015, there were more than 300 code metrics proposed or analyzed; code complexity, in particular, was one of the most common category observed in this study. This high number of complexity metrics can challenge software engineers.

Other studies focused on understanding developers' challenges in program comprehension tasks. 
Santos and Gerosa~\cite{Santos:ICPC:2018} asked 62 survey participants to grade well-known coding practices. The authors found that 7 (out of the 11 tested practices) tend to increase readability. Wiese at al.~\cite{Wiese:ICPC:2021} studied how students (mis)understand code snippets containing multiple Boolean expressions. The authors found that students tend to misunderstand conditions when several Boolean expressions are conjoined; if the expressions are separated in individual \texttt{if} statements, students tend to understand better.

More recently, some researchers are employing eye-tracking tools to trace participants' eye movements and thus figure out the parts of the code on which they focus their effort. For instance, Nahla and colleagues~\cite{Nahla:ICSE:2019} found that developers tend to focus most of their time and attention on the body of the methods instead of the methods' signature, which contrast with previous studies. They also observed that experienced developers tend to revisit the body of the methods more often as the method size increases.
Other studies used fMRI technologies to observe how programmers comprehend short code snippets. For instance, Peitek and colleagues~\cite{peitek2021program} conducted a study using fMRI with 19 participants. Among the findings, the authors found that using simple metrics, such as LOC and McCabe, have good performance when predicting cognitive load.

Finally, although there are some early works on CDD (e.g., \cite{CDD:ESEM:2022,Pinto:ICEIS:2022,Souza:ICSME:2020}), none of these works assessed the developers' perception in dealing with CDD in practice. Our work extends these works by bringing a unique perception of the use of CDD from the trenches.



\section{Conclusion}

CDD is a coding design technique that aims to reduce the cognitive load that developers may face when reading code by limiting the number of programming constructs that could be used in a given code unit.
Here we shared our experience in using CDD to build a Java-based online learning management tool at Zup Innovation, a Brazilian tech company. By combining our observations from the developers' work with data and metadata from the repositories, we could curate a set of findings on the CDD practice. We highlight that:

\begin{enumerate}
    \item CDD could help software development teams to keep the size of the code units reasonably small, potentially helping to identify god classes earlier;
    \item by having small code units, CDD also impacts testing practices; according to an engineer ``since CDD leads to small code units, these code units are also easier to tests''; 
    \item CDD brings more awareness for refactoring operations; instead of refactoring based on their intuitions, developers following CDD practices have now the evidence of when refactorings are really needed.
\end{enumerate}



\vspace{0.2cm}
\noindent
\textbf{Future work.}
First, we plan to use CDD in the other services that compose \handora. This was not possible during this research, because the team working on them was not yet used to CDD. For instance, we are interested in understanding how CDD could help the development of the machine learning service (S6) and the frontend service (S5). Moreover, we did not use CDD for writing testing code. We understand that by limiting the size of production classes and methods, the testing classes may also be positively impacted. In the future, we plan to conduct experiments to understand to what extent it makes sense to use CDD in designing testing code.

Second, we observed that CDD helped the software development team to keep the classes under the defined limit. However, as a potential side effect, these small classes might have a more significant number of relationships. We then plan to understand whether these small classes require more navigation from the developer, which in turn could hinder program comprehension. 

Finally, we plan to build production-ready static analysis tools that support CDD. Using this tool, we plan to perform controlled experiments to understand to what extent the tool could help developers find classes worth refactoring according to CDD principles. We also plan to create a bot that computes the CDD metrics during the CI/CD process. We believe this bot could ease the CDD adoption in software-producing teams.

\vspace{0.2cm}
\noindent
\emph{Acknowledgments.} We thank the maintainers for reviewing our patches; and the reviewers of SCAM and JSS for their helpful comments.
This work was partially supported by CNPq (\#406308/2016-0 and \#309032/2019-9);
and by the Swiss National Science Foundation (SNSF) grant \emph{Hi-Fi} (\#200021\_182060).

\balance
\bibliographystyle{IEEEtran}
\bibliography{main}

\end{document}